\documentclass[aps,prl,superscriptaddress,twocolumn,longbibliography]{revtex4-2}
\usepackage{bbm}
\usepackage{mathtools}
\usepackage{graphicx}
\usepackage{dcolumn}
\usepackage{bm}
\usepackage{subfigure}
\usepackage{amsmath}
\usepackage{amssymb}
\usepackage{feynmf}
\usepackage{hyperref}
\usepackage{color}

\usepackage{times}
\begin{document}

\title{Unveiling the multiband metallic nature of the normal state in nickelate La$_3$Ni$_2$O$_7$}

\author{Bowen Chen}
\affiliation{Center for Neutron Science and Technology, Guangdong Provincial Key Laboratory of Magnetoelectric Physics and Devices, School of Physics, Sun Yat-Sen University, Guangzhou, 510275, China}
\affiliation{State Key Laboratory of Optoelectronic Materials and Technologies, Sun Yat-Sen University, Guangzhou, Guangdong 510275, China}

\author{Hengyuan Zhang}
\affiliation{Center for Neutron Science and Technology, Guangdong Provincial Key Laboratory of Magnetoelectric Physics and Devices, School of Physics, Sun Yat-Sen University, Guangzhou, 510275, China}

\author{Jingyuan Li}
\affiliation{Center for Neutron Science and Technology, Guangdong Provincial Key Laboratory of Magnetoelectric Physics and Devices, School of Physics, Sun Yat-Sen University, Guangzhou, 510275, China}

\author{Deyuan Hu}
\affiliation{Center for Neutron Science and Technology, Guangdong Provincial Key Laboratory of Magnetoelectric Physics and Devices, School of Physics, Sun Yat-Sen University, Guangzhou, 510275, China}

\author{Mengwu Huo}
\affiliation{Center for Neutron Science and Technology, Guangdong Provincial Key Laboratory of Magnetoelectric Physics and Devices, School of Physics, Sun Yat-Sen University, Guangzhou, 510275, China}

\author{Shuyang Wang}
\affiliation{Anhui Key Laboratory of Low-Energy Quantum Materials and Devices,High Magnetic Field Laboratory, HFIPS, Chinese Academy of Sciences, Hefei, Anhui 230031, China}

\author{Chuanying Xi}
\affiliation{Anhui Key Laboratory of Low-Energy Quantum Materials and Devices,High Magnetic Field Laboratory, HFIPS, Chinese Academy of Sciences, Hefei, Anhui 230031, China}

\author{Zhaosheng Wang}
\email{Corresponding author: zswang@hmfl.ac.cn}
\affiliation{Anhui Key Laboratory of Low-Energy Quantum Materials and Devices,High Magnetic Field Laboratory, HFIPS, Chinese Academy of Sciences, Hefei, Anhui 230031, China}

\author{Hualei Sun}
\email{Corresponding author: sunhlei@mail.sysu.edu.cn}
\affiliation{Center for Neutron Science and Technology, Guangdong Provincial Key Laboratory of Magnetoelectric Physics and Devices, School of Physics, Sun Yat-Sen University, Guangzhou, 510275, China}

\author{Meng Wang}
\email{Corresponding author: wangmeng5@mail.sysu.edu.cn}
\affiliation{Center for Neutron Science and Technology, Guangdong Provincial Key Laboratory of Magnetoelectric Physics and Devices, School of Physics, Sun Yat-Sen University, Guangzhou, 510275, China}

\author{Bing Shen}
\email{Corresponding author: shenbing@mail.sysu.edu.cn }
\affiliation{Center for Neutron Science and Technology, Guangdong Provincial Key Laboratory of Magnetoelectric Physics and Devices, School of Physics, Sun Yat-Sen University, Guangzhou, 510275, China}
\affiliation{State Key Laboratory of Optoelectronic Materials and Technologies, Sun Yat-Sen University, Guangzhou, Guangdong 510275, China}

\begin{abstract}
The discovery of unconventional superconductivity around 80 K in perovskite nickelates under high pressure has furnished a new platform to explore high-temperature unconventional superconductivity in addition to cuprates.  Understanding the normal state of nickelate superconductors is crucial to uncovering the origin of this unconventional superconductivity and gaining further insight into its underlying mechanism. In this study, we systemically studied the transport properties of La$_3$Ni$_2$O$_7$ by tuning the pressure under high magnetic fields. Magnetoresistance (MR) consistently exhibits a quasi-quadratic dependence on the magnetic field across all measured pressures and temperatures. Increased pressure enhances the metallicity of the system and leads to a monotonic increase in MR, which follows the extended Kohler’s rule. These results suggest that the normal state of La$_3$Ni$_2$O$_7$ to be a multiband metallic nature. 

\end{abstract}
\date{\today}
\maketitle

\section{I. INTRODUCTION}
The discovery of superconductivity at 80 K under high pressure in bulk nickelate marks a groundbreaking achievement for the observation of a new type of liquid-nitrogen-temperature strongly correlated materials alongside cuprates\cite{Sun2023,Hou2023,Zhang2024}. Experimental and theoretical studies about Ruddlesden-Popper (RP) phases of La$_{n+1}$Ni$_n$O$_{3n+1}$ family (n=1-3), particularly for n=2 and 3, have expanded rapidly and garnered significant attention in condensed matter physics\cite{BILA_TWO_MODE,Wang2024,SUPE_4310,SUPE_LA4310_2,ZHANG2024147,INVE_OF_KEY_ISSUE}. The bilayer compound La$_3$Ni$_2$O$_7$ and  La$_2$PrNi$_2$O$_7$ under high pressure can demonstrate bulk superconductivity with  onset superconducting transition temperatures ($T_c$) exceeding the liquid-nitrogen temperature\cite{Sun2023,Wang2024}. The trilayer compound La$_4$Ni$_3$O$_{10}$  exhibits bulk superconductivity with a $T_c$ of 30 K under high pressure and a superconducting volume fraction of nearly 80 $\%$ \cite{SUPE_4310,SUPE_LA4310_2}. Recent studies have also reported signatures of superconductivity in pressurized trilayer-nickelate Pr$_4$Ni$_3$O$_{10}$ \cite{Huang_2024_PrNiO}. The exploration for higher superconductivity is still in full swing in more nickelate compounds. 

For bilayer nickelate compounds with high $T_c$ comparable to cuprates, high pressure drives a rich phase diagram\cite{Sun2023,Hou2023,Zhang2024,Geisler2024,structure_2,Li_2024,STRU_PHAS_TRAN}.  At ambient pressure, various experiments such as resonant inelastic X-ray scattering (RIXS) \cite{Chen2024}, nuclear magnetic resonance spectroscopy measurements(NMR)\cite{NMR1,2024arXiv240203952D}, muon spin relaxation measurements($\mu$sR)\cite{miuSR_spin_density_wave,miusR_2}  and etc\cite{Liu2022}, suggest the presence of a ($\pi$/2,$\pi$/2,$\pi$) spin-density-wave order (SDW) below 150 K. Infrared spectroscopy measurements\cite{Liu2024} have revealed a change density wave (CDW) emerging around 115 K. However, recent neutron scattering results have reported the absence of  long-range orders observed above 10 K although strong antiferromagnetic (AF) spin fluctuations are observed\cite{XIE20243221}. Additionally,  angle-resolved photoemission spectroscopy (ARPES) measurements recognize this ``parent" compound as a metal with two types of Fermi pockets\cite{Yang2024,BILA_TWO_MODE}. Furthermore, properties such as resistivity and superconductivity in nickelates are known to be highly sensitive to oxygen defects and their inhomogeneities\cite{Dong2024,miuSR_spin_density_wave,Gao2024,oxygen_deficiencies_1,oxygen_deficiencies_2}. Under high pressure, the crystal structure of La$_3$Ni$_2$O$_7$ undergoes a phase transition from the orthorhombic symmetry $Amam$ to $I4/mmm$ at approximately 14 GPa, coinciding with the emergence of superconductivity\cite{Geisler2024,structure_2,Li_2024}. This structural change raises the question of whether the ambient-pressure phase can truly be considered a ``parent phase" of the superconducting state.  The observed abnormal transport behaviors and theoretically predicted exotic orders further complicate and deepen the mystery surrounding this unconventional superconductivity\cite{Wang_2024,s_wave,CORR_ELEC_STRU,ELEC_CORR_AND_SUPE_INST,ELEC_DIME_ORBI,CORR_ELEC_STRU_ORBI,BILA_TWO_MODE}. 

Understanding the normal state of nickelate superconductors is crucial to reveal the origin of this superconductivity and gaining deeper insight into the unconventional high-temperature superconducting mechanism\cite{EMER_SUPE_UNIV,CUPR,BiSrLaCuO,LINE_NdSRNIO}. Systematic investigations for the normal state of nickelates, particularly under pressure, are both urgent and necessary \cite{Liu2024,FLATB}. However, due to the experimental challenges associated with high-pressure measurements, relevant studies remain limited. In this study, we performed systematic transport measurements of La$_3$Ni$_2$O$_7$ under high pressures with high magnetic fields. Across the entire pressure and temperature range studied,  all magnetoresistance (${\rm MR}\equiv(R(\mu_0H)-R(0))/R(0)$, where $R(0)$ is the resistance at at zero field) demonstrate a quasi-quadratic  dependence on the applied field. As pressure increased, the system became progressively more metallic, leading to a monotonic increase in MR.  Using an extended Kohl's rule, the MR is well scaled, and the extracted thermal factor $n_T$ exhibits strong temperature dependence. The results suggest a multiband metallic normal state for La$_3$Ni$_2$O$_7$ distinguishing it from  cuprates.

\section{II. EXPERIMENTAL DETAILS}
The single crystals of La$_3$Ni$_2$O$_7$ were grown by using a vertical optical-image floating-zone furnace under an oxygen pressure of 15 bar, utilizing a 5-kW Xenon arc lamp (100-bar Model HKZ, SciDre)\cite{Sun2023}. Electrical resistance measurements of single crystals of La$_3$Ni$_2$O$_7$ were performed by using the standard four-probe
method. High pressure was generated with screw-pressure-type DAC 
made of nonmagnetic Be-Cu alloy. Diamond anvils with a 300 $\mu$m 
culet was used, and the corresponding sample chamber with a 
diameter of 100 $\mu$m was made in an insulating gasket achieved by 
cubic boron nitride and epoxy mixture. A single crystal with a 
dimension of 100$\times$100$\times$40 $\mu$m$^3$ was loaded with KBr 
powders as the pressure-transmitting medium. Pressure calibrations were carried using the ruby fluorescence shift at room temperature for all experiments. Initial electrical measurements were performed on a physical property measurement system (PPMS, Quantum Design), which provides extreme environments with temperatures ranging from 2 K to 300 K and magnetic fields up to 14 T. Subsequently, high-field MR measurements were performed on the Steady High Magnetic Field Facilities (WM5), High Magnetic Field Laboratory, Chinese Academy of Sciences. A LakeShore model 370 AC resistance bridge was used for data collection.

\section{III. RESULTS AND DISCUSSIONS}
\begin{figure}
    \centering
    \includegraphics[width=3.5in]{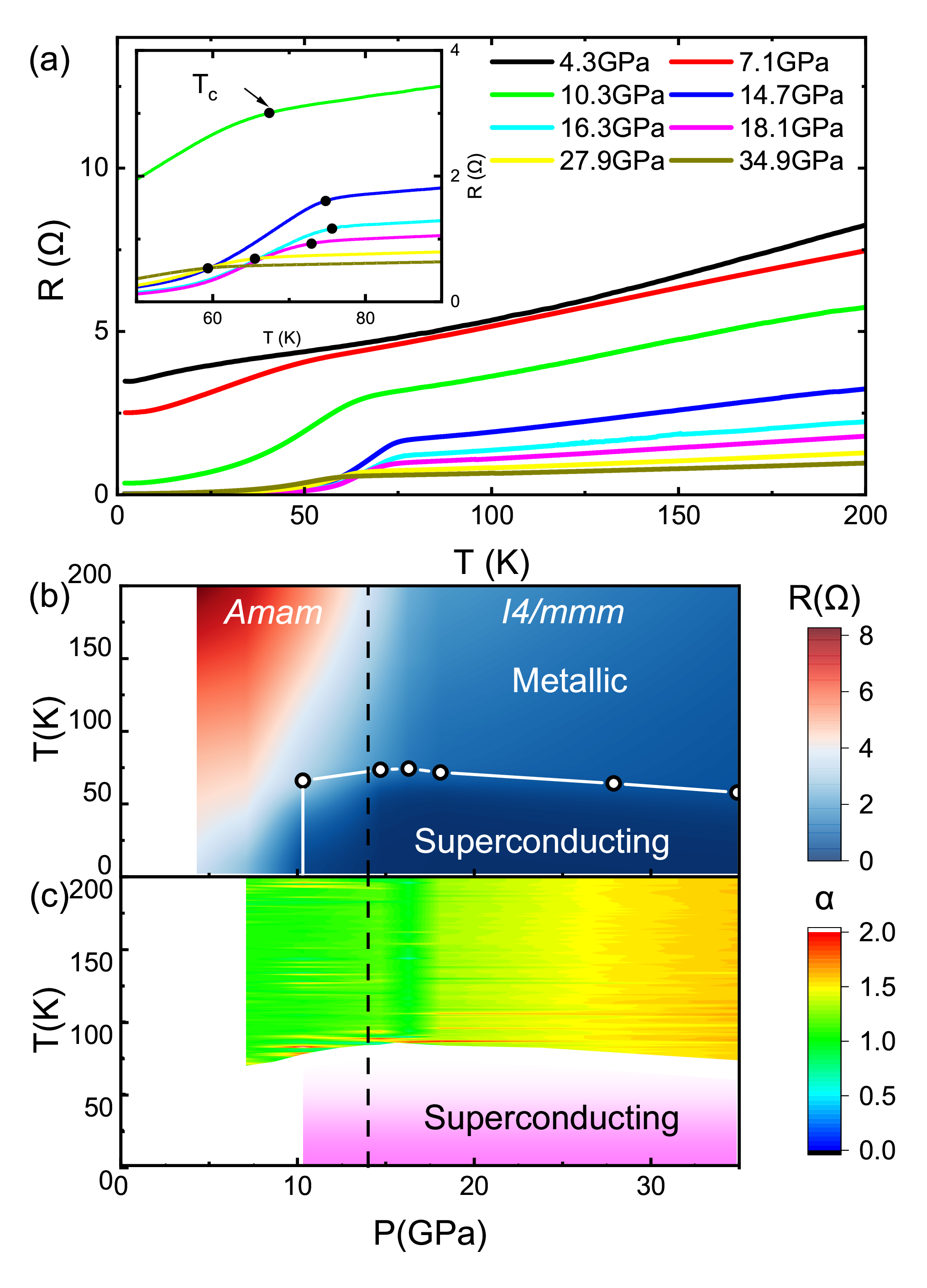}
    \caption{ (a) Temperature dependent resistance $R(T)$ of La$_3$Ni$_2$O$_7$ at various pressures.  (b) The phase diagram of La$_3$Ni$_2$O$_7$ acquired from the $R(T)$ feature.  The white circle denotes the superconducting transition temperature at various pressures. The black dash line mark the boundary of different crystal structure according to former reports\cite{Li_2024}. The color represents the $R(T)$ value. (c) The color map of fitting parameter $\alpha$ for normal state resistance data by using a function $R_{xx}(T)=R_0+AT^\alpha$. .}
    \label{fig:FIG. 1.}
\end{figure}

Figure 1 shows the temperature dependence of the resistance ($R(T)$) for a single crystal of La$_3$Ni$_2$O$_7$ with tuning the pressure range from 4.5 GPa to 34.9 GPa. In this study, all our measurements under high magnetic fields were performed on the same crystal and during a single increased pressure run. As pressure is applied, the system undergoes a structure transition from $Amam$ to $I4/mmm$ phase\cite{Geisler2024,structure_2,Li_2024}. Superconductivity first appears at 10.3 GPa with a critical temperature  $T_c$=67 K, and reaches its maximum $T_c$=75 K at 16.3 GPa. For ease of discussion, we designate the pressure range where superconductivity emerges as the ``under" region and the point with the highest $T_c$   as the ``optimal" region.  Beyond 18.1 GPa, increasing pressure suppresses superconductivity, and the critical temperature decreases to 59 K, which we define as the ``over" region. The systematic evolution of superconductivity and its phase diagram are shown in Fig. 1 (b) consistent with previous reports\cite{Sun2023,Hou2023,Zhang2024,Li_2024}. To analyze the normal state, the $R(T)$ curves at various pressures can be fitted by using the following function:
    \begin{equation}
        R_{xx}(T)=R_0+AT^\alpha,
    \end{equation} 
where $R_0$ is the residual resistance, and $\alpha$ is the temperature exponent. For obtaining the evolution of $\alpha$, the function is fitted with a sliding temperature windows with a width $\Delta T=6$ K at different pressures, as illustrated in Fig. 1 (c). Above $T_c$, the temperature exponent $\alpha$ remains around 1 in both the ``under" and ``optimal" regions. For the ``over" region, $\alpha$ increases monotonically, reaching to 1.6 as pressure increases the pressure to 34.5 GPa. These trends suggest that the normal state in the ``under" region, previously identified as a ``strange metal" in earlier studies, becomes increasingly metallic with higher pressure.    

\begin{figure*}
    \centering
    \includegraphics[width=7in]{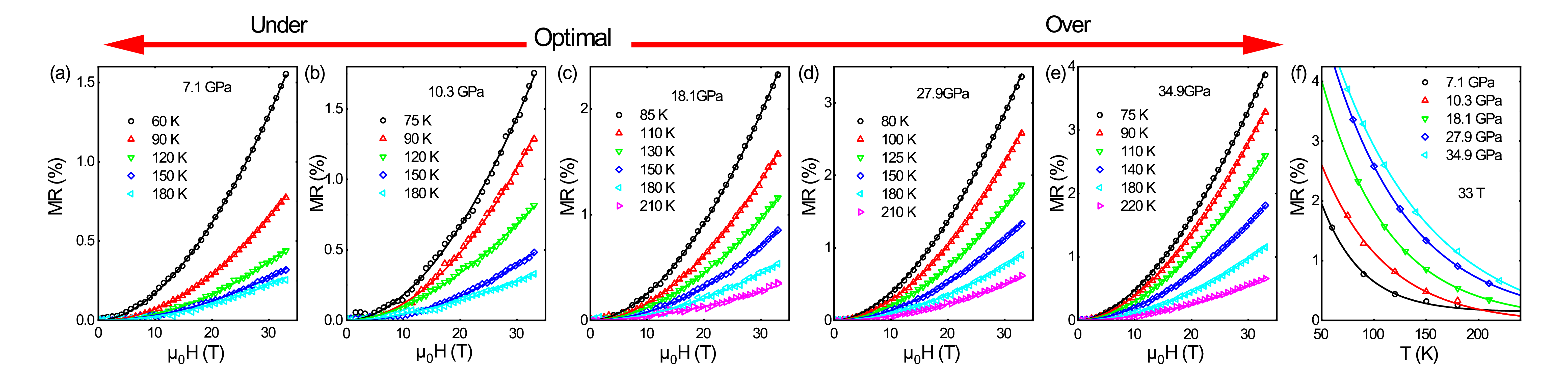}
    \caption{ (a)-(e)  Field-dependent MR of La$_3$Ni$_2$O$_7$ at different various temperatures and pressures. The open symbols represent the experimental data and the solid lines represent the fitting curves by using the formula ${\rm MR}(\mu_0H)=a(\mu_0H)^n$. (f) The temperature-dependent MR of La$_3$Ni$_2$O$_7$ with $\mu_0H$=33 T at various pressures. The solid symbols represent the experimental data and the solid lines represent fitted by using exponential functions. }
    \label{fig:FIG. 2.}
\end{figure*}

To further investigate the normal state for La$_3$Ni$_2$O$_7$, systematic MR measurements under high magnetic fields were conducted as shown in 
Figs.2 (a)-(e). Above $T_c$,  the MR decreases monotonically with increasing temperature. Although various strange orders such as CDW and SDW have been proposed, the observed field-dependent and pressure-dependent MR (MR($\mu_0H$) and MR($T$)) exhibit smooth and continuous evolutions, as shown in Figs.2 (a)-(f). To further investigate the field dependence of MR, MR($\mu_0H$) curves at various pressures are fitted by using the following function:
  \begin{equation}
      {\rm MR}=a(\mu_0H)^n,
  \end{equation}
where $a$ is the transverse MR coefficient, and $n$ is the field exponent.  The extracted field exponents range from 1.75 to 2.1 across the entire temperature and pressure range, suggesting a typical quadratic field dependence which can be understood by the semiclassical model\cite{Olsen1962ElectronTI}.

\begin{figure*}
    \centering
    \includegraphics[width=7in]{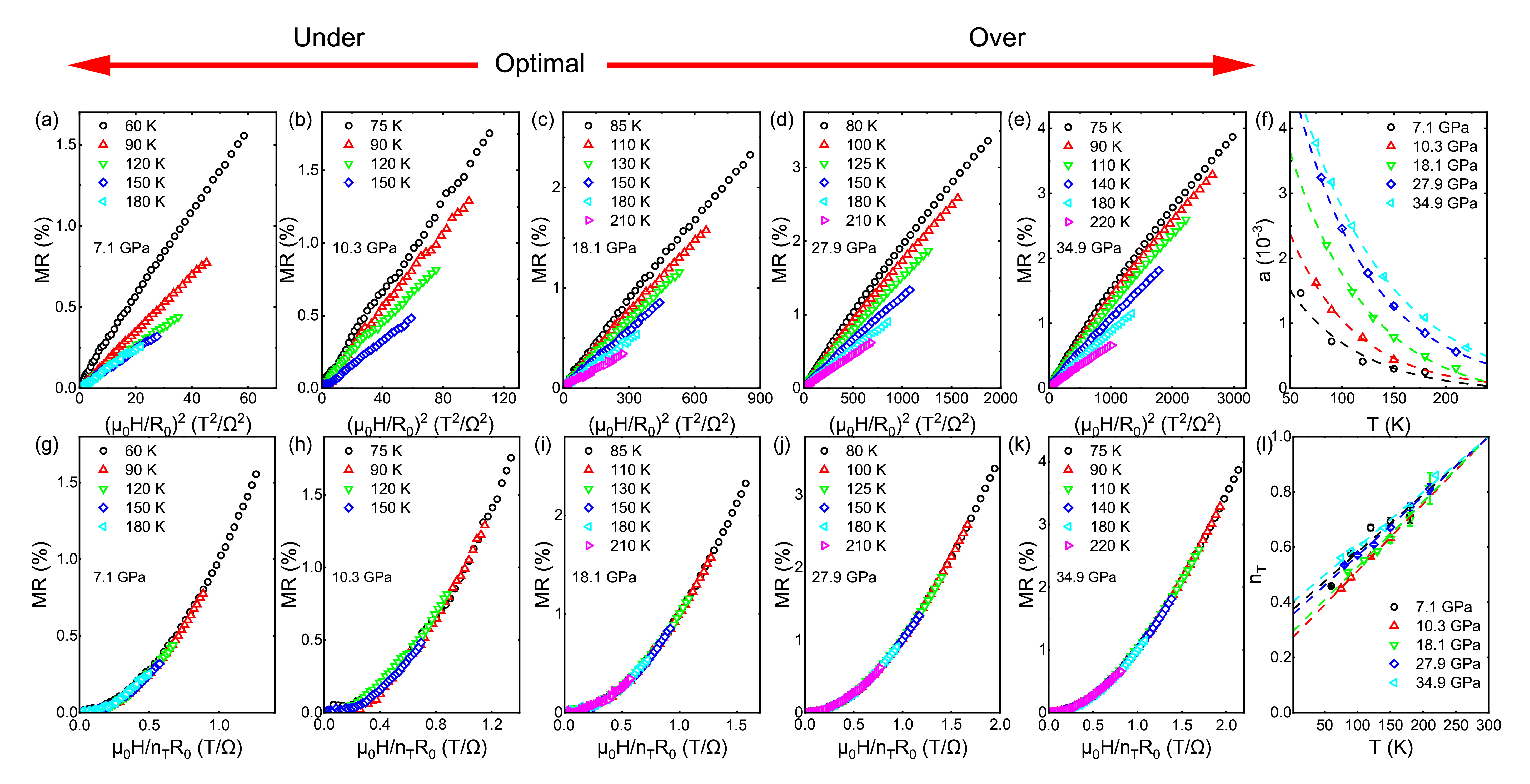}
    \caption{ (a)-(e)  Kohler plots of La$_3$Ni$_2$O$_7$ at various pressures. (f) The temperature-dependent coefficients of La$_3$Ni$_2$O$_7$ at various pressures. The open symbols represent the experimental data and the dashed lines represent the fitting curves by using exponential functions. (g)-(k) Extended Kohler plots of  La$_3$Ni$_2$O$_7$ at various pressures. (i) The temperature-dependent thermal factors $n_T$ of La$_3$Ni$_2$O$_7$ at various pressures. The open symbols represent the experimental data. }
    \label{fig:FIG. 3.}
\end{figure*}

For a conventional metal in the Fermi-liquid state, the orbital motion of carriers can be scaled by the product of the cyclotron frequency $\omega_c$ and scattering time $\tau$. And the $\omega_c$ and $\tau$ are proportional to field $\mu_0H$ and inverse resistance\cite{EVOL_FROM_NONF}. Consequently, MR should obey Kohler's rule expressed as ${\rm MR}=f(\mu_0H/R(0))$. This relation arises because the magnetic field enters Boltzmann’s equation in the combination
$\mu_0H\tau$ and that R$_0$ is proportional to the scattering rate
$1/\tau$. In the weak-field limit, most simple metals exhibit a quadratic field dependence of the MR, described by MR$\sim$$(\mu_0H)^2\tau^2$. Therefore, a plot of MR versus $(\mu_0H/R(0))^2$ is expected to collapse to a single temperature-independent curve if the carrier density remains constant. Additionally, the temperature dependence of the scattering rate should be roughly uniform across the Fermi surface or various Fermi pockets  which is most easily satisfied when there is only a single, temperature-dependent scattering rate.\cite{IN_PLAN_MAGN}.  Although the MR of La$_3$Ni$_2$O$_7$ exhibits the quasi-quadratic field dependence, it still violates Kohler's rule as shown in Figs.3 (a)-(e) with the $e^{kT}$ temperature dependence transverse MR coefficients shown in Fig.3 (f).

This violation of Kohler's rule is reminiscent of the similar behavior observed in both the Psudogap and strange-metal regimes of La$_{2-x}$Sr$_x$CuO$_4$ 
and YBa$_2$Cu$_3$O$_6$\cite{VIOL_OF_YBCO_LSCO}, where the unusual MR behavior has been attributed to the anomalous temperature dependence of the Hall angle\cite{VIOL_OF_YBCO_LSCO}. Unlike these cuprates, which have a single type of Fermi sheet, La$_3$Ni$_2$O$_7$ features a more complex Fermi topology, consisting of at least two types of Fermi sheets: an electron-like $\alpha$ sheet and a hole-like $\beta$ sheet \cite{BILA_TWO_MODE}. Under high pressure, the system undergoes a structure transition from the orthorhombic phase($Amam$) to another tetragonal phase with a space group($I4/mmm$). Some theoretical studies suggest that under high pressure the $\gamma$ bands derived from the Ni-3$d$ orbital cross the Fermi level\cite{Sun2023},  while others propose that only $\alpha$ and $\beta$ sheets cross, accompanied by Fermi topology evolution\cite{PhysRevB.110.205122}. These studies suggest that the prominent multiband electron structure can host diverse temperature-dependent scatterings or carriers from different bands, which results in the violation of Kohler's rule. To account for this, we adopt an extended Kohler’s rule, incorporating a thermal factor $n_T$ to scale our MR data, as shown in Figs.3(g)-(k)\cite{EXTE_KOHL}. The MR can be scaled by the following function:
\begin{equation}
      {\rm MR}=f(\frac{\mu_0H}{n_TR(0)}),
  \end{equation}
where $R(0)$ is the resistance at 0 T, and $n_T$ is the thermal factor. All our MR data can be well scaled onto one curve at each pressure.  
The extended Kohler's rule takes the change in carrier density and mobility caused by the thermal excitation into considerations. Here the thermal factor $n_T$ is expressed as\cite{EXTE_KOHL} 
\begin{equation}
n_T=e[\sum_i(n_i\mu_i)]^{3/2}/[\sum_i(n_i\mu_i^3)]^{1/2}
\end{equation}
where $n_i$ and $\mu_i$ are the $i^{th}$-band's carrier density and mobility respectively.  The thermal factor $n_T$ describes the relative change induced
by thermal excitation for the effective carrier density, which is associated with the Fermi level and the dispersion relation. When the effective carrier density is largely temperature-independent, Equation (3) with $n_T$=1 is denoted as the original Kohler's rule. 
   
Considering the two types of Fermi sheets observed in La$_3$Ni$_2$O$_7$\cite{BILA_TWO_MODE,EXTE_KOHL}, $n_T$ can be expressed as:
\begin{equation}
n_T=\frac{e(n_h\mu_h-n_e\mu_e)^{3/2}}{(n_h\mu_h^3-n_e\mu_e^3)^{1/2}}
\end{equation}
where $n_h$ and $n_e$ represent the carrier density for hole and electron, respectively,  and $\mu_h$ and $\mu_e$ tare their corresponding mobilities. The $n_T$ values obtained from the extended Kohler’s rule scaling at various pressures are shown in Fig. 3(i), demonstrating a smooth linear temperature dependence. Unlike pnicitides BaFe$_2$(As$_{1-x}$P$_x$)$_2$ ($x$=0-1) and topological TaAs\cite{EXTE_KOHL}, no anomalies such as discontinuities or saturation are observed in  $n_T(T)$. 

\begin{figure}
    \centering
    \includegraphics[width=3.5in]{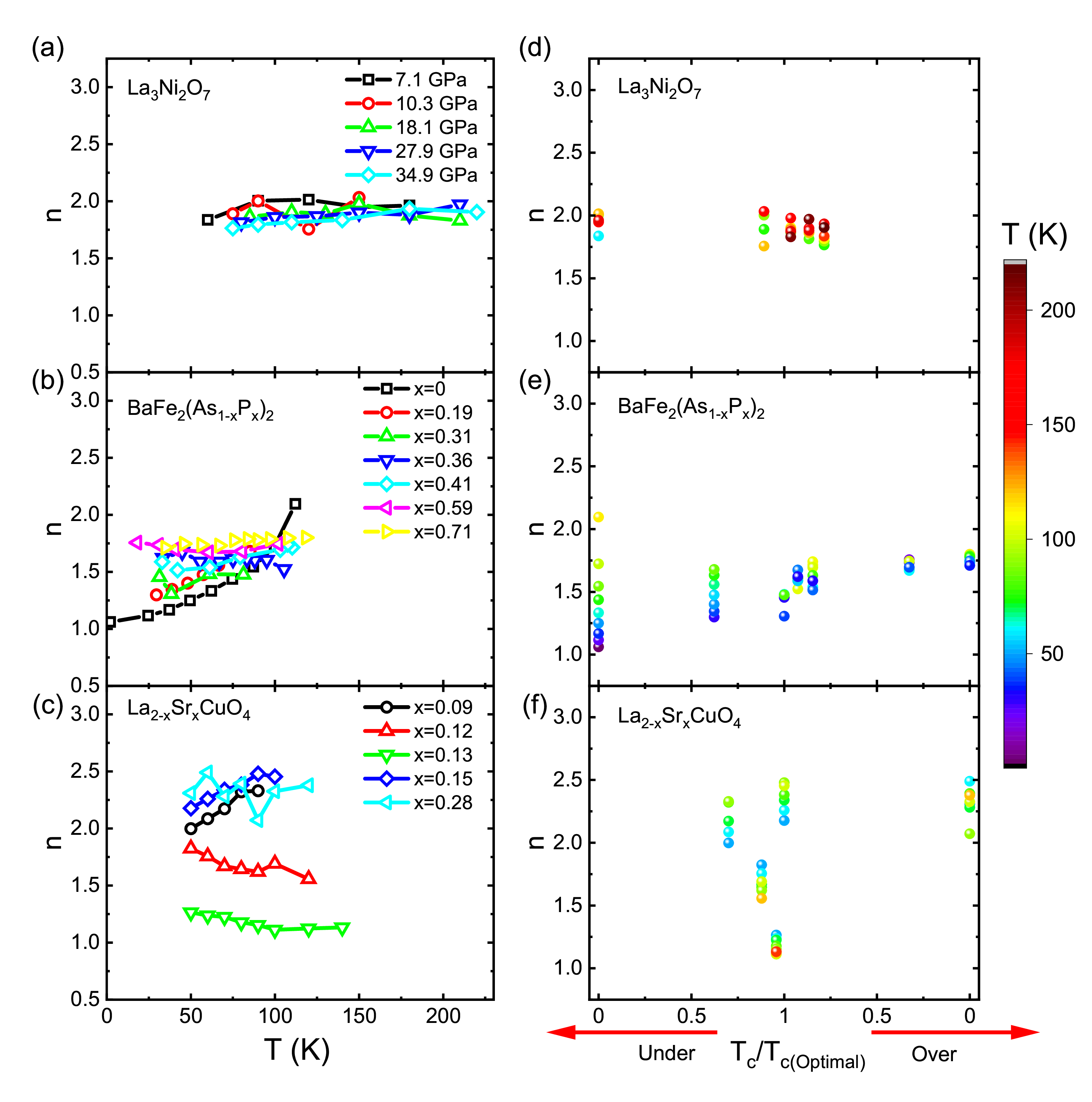}
    \caption{ (a) Temperature-dependent coefficient $n$ of La$_3$Ni$_2$O$_7$ at various pressures. (b) Temperature-dependent coefficient $n$ of BaFe$_2$(As$_{1-x}$P$_x$)$_2$ with different doping levels. The data of $x$=0 and $x$=0.19 are taken from Ref\cite{MAGN_SCAL_BFAP}, and the data of $x$=0.31-0.71 are taken from Ref\cite{SCAL_BETW_MAGN}. (c) Temperature-dependent coefficient $n$ of La$_{2-x}$Sr$_x$CuO$_4$ with different doping levels.The data are taken from Ref\cite{INPL_OUPL_LSCU}. (d)-(f) The coefficient $n$ versus reduced superconducting transition temperature $T_c/T_{c(Optimal)}$ of  La$_3$Ni$_2$O$_7$, BaFe$_2$(As$_{1-x}$P$_x$)$_2$ and La$_{2-x}$Sr$_x$CuO$_4$, respectively.}
    \label{fig:FIG. 4.}
\end{figure}

Fig.4 summarizes the field exponents $n$ from Equation (2) for several typical unconventional high-temperature superconductors.  For La$_3$Ni$_2$O$_7$, $n$ remains consistently around 1.9, indicating the quasi-quadratic field dependence for the MR, independent of temperature and pressure. In contrast, the multiband pnictides BaFe$_2$(As$_{1-x}$P$_x$)$_2$ ($x$=0-1), which exhibits a rich phase diagram featuring  a interplay of  of superconductivity, SDW, nematic order, and a hidden quantum critical point (QCP)\cite{EVOL_FROM_NONF,TRAN_NEAR,ELEC_NEMA}, demonstrate strong temperature and chemical pressure dependence of $n$. In the parent compound BaFe$_2$As$_2$, the MR changes from a quadratic field dependence to a linear field dependence as the SDW develops upon cooling. Additionally, chemical pressure tuning induces a crossover in the MR field dependence, transitioning from Fermi liquid to non-Fermi liquid behavior. For cuprates such as La$_{2-x}$Sr$_x$CuO$_4$, the MR demonstrates minimal temperature dependence but varies significantly  across different doping levels, despite only minor changes in $T_c$\cite{INPL_OUPL_LSCU}. These strongly correlated ``bad metals" deviate markedly from the quadratic field dependence expected for a typical Fermi liquid, reflecting their unconventional nature.
 
In cuprates and pnictides superconductors, the MR is highly sensitive to the
emergence of exotic orders and non-Fermi liquid states. The significant 
violations of Kohler's rule and Fermi-liquid transport behavior in 
cuprates highlight the unconventional  nature of their normal state.  In the pnictides BaFe$_2$(As$_{1-x}$P$_x$)$_2$, a extended Kohler’s rule well describes this multiband 
system, revealing a temperature dependent thermal factor $n_T(T)=n_0+aTe^{-\Delta /k_BT}$. The observed abnormal kink in $n_T(T)$ near the chemical pressure of QCP suggests the emergence of a possible pseudogap\cite{EXTE_KOHL,INTE_OF_LSCO,QUAS_RELA_YBCO}. The MR in this system exhibits an unusual linear dependence on both temperature and field, which can be scaled using the function $\sqrt{(\alpha k_BT)^2+(\gamma \mu_B \mu_0H)^2}$, indicating a Fermi liquid to non-Fermi liquid crossover. In sharp contrast to those high-temperature superconductors, La$_3$Ni$_2$O$_7$ exhibits  MR with consistent temperature and field dependence across nearly all temperatures and pressures. While the extended Kohler’s rule successfully scales all MR data at various pressures, the revealed thermal factor $n_T$ displays a smooth, linear temperature dependence without any anomalous features.  These straightforward MR behaviors in La$_3$Ni$_2$O$_7$ align with a multiband metallic system described by the semiclassical model. In the ``strange" metal region of La$_3$Ni$_2$O$_7$, the MR appears surprisingly ``normal" compared to that of cuprates and pnictides, despite this bilayer compound achieving much higher superconducting transition temperatures than trilayer counterparts. Recent theoretical studies suggest that local interlayer spin interactions play a significant role in the high-temperature superconductivity of this material, emphasizing the importance of interlayer effects\cite{HALF_FILL_BILA,INTE_VALE_BOND}. Consequently, the transport behaviors within the $ab$ plane dominated by intralayer's carriers and scatterings resemble those of a conventional multiband metal. Investigations of transport along the $c$-axis are expected to reveal critical interlayer information. Further relevant studies of this bilayer structure may uncover key insights and potential pathways for the development of new high-temperature superconductors.

\section{IV. SUMMARY}
The MR of La$_3$Ni$_2$O$_7$ under high magnetic fields was systematically studied by varying pressure across a range of temperatures. It increases monotonically with pressure and  exhibits a consistent quasi-quadratic field dependence at all pressures and temperatures. Using an extended Kohler’s rule, the MR can be well scaled under all pressures.  Our results suggest the transport behaviors of the normal states in nickelate superconductors can be roughly understood by a multiband metal model. Notably, the thermal factor $n_T$ exhibits a linear temperature dependence without anomalies, in stark contrast to the behavior observed in cuprate and pnictide superconductors.

\section{ACKNOWLEDGMENTS}
This project is supported by the National Key
R\&D Program of China (Grant Nos. 2023YFF0718400 and
2023YFA1406500), the open research fund of Natural Science Foundation of Guangdong Province (Grant No. 2022A1515010035), the National Natural Science Foundation of China (Grant
Nos. U2130101, 92165204, 12425404 and 12174454), Guangzhou Basic and Applied Basic Research Foundation(Grant No. 202201011798), Songshan Lake Materials Laboratory (Grant Nos. 2021SLABFN11 and 2024B1515020040),the Open Project of Guangdong Provincial Key Laboratory of Magnetoelectric Physics and Devices (Grant No. 2022B1212010008), the Open Project
of Key Laboratory of Optoelectronic Materials and Technologies (Grant No. OEMT-2023-ZTS-01). The authors thank Dr. Long Jiang from the Instrumental Analysis \& Research Center of Sun Yat-Sen University for single-crystal X-ray diffraction measurements and structural analysis.
Part of this work was performed on the Steady High Magnetic Field Facilities, High Magnetic Field Laboratory, Chinese Academy of Sciences, and supported by the High Magnetic Field Laboratory of Anhui Province.

\end{document}